\documentclass{jpsj3}
\usepackage{txfonts}
\usepackage[mathlines, pagewise,displaymath]{lineno}
\newcommand{\argmax}{\mathop{\rm arg~max}\limits}
\newcommand{\argmin}{\mathop{\rm arg~min}\limits}

\title{Simultaneous Estimation of Noise Variance and Number of Peaks in Bayesian Spectral Deconvolution}

\author{Satoru Tokuda$^1$, Kenji Nagata$^2$$^3$, and Masato Okada$^1$$^4$$^5$\thanks{okada@k.u-tokyo.ac.jp}}
\inst{$^1$Department of Complexity Science and Engineering, The University of Tokyo, Kashiwa, Chiba 277-8561, Japan \\
$^2$Artificial Intelligence Research Center, AIST, Koto, Tokyo 135-0064, Japan \\
$^3$PRESTO, Japan Science and Technology Agency, Kawaguchi, Saitama 332-0012, Japan \\
$^4$RIKEN Brain Science Institute, Wako, Saitama 351-0198, Japan \\
$^5$Center for Materials Research by Information Integration, National Institute for Materials Science,
Tsukuba, Ibaraki 305-0047, Japan}

\abst{The heuristic identification of peaks from noisy complex spectra often leads to misunderstanding of the physical and chemical properties of matter. In this paper, we propose a framework based on Bayesian inference, which enables us to separate multipeak spectra into single peaks statistically and consists of two steps. The first step is estimating both the noise variance and the number of peaks as hyperparameters based on Bayes free energy, which generally is not analytically tractable. The second step is fitting the parameters of each peak function to the given spectrum by calculating the posterior density, which has a problem of local minima and saddles since multipeak models are nonlinear and hierarchical.
Our framework enables the escape from local minima or saddles by using the exchange Monte Carlo method and calculates Bayes free energy via the multiple histogram method. 
We discuss a simulation demonstrating how efficient our framework is and show that estimating both the noise variance and the number of peaks prevents overfitting, overpenalizing, and misunderstanding the precision of parameter estimation.
}

\begin{document}
\maketitle
\nolinenumbers

\section{Introduction}
Spectroscopy is at the heart of all sciences concerned with matter and energy.
An electromagnetic spectrum indicates the electronic states and the kinetics of atoms.
The quantum nature of spectra allows them to be approximately reduced to the sum of unimodal peaks (such as Lorentzian peaks, Gaussian peaks, and their convolutions), whose centers are the energy levels from the semiclassical viewpoint \cite{tkachenko2006optical}. 
The peak intensity is proportional to both the population density of the atoms or molecules and their transition probabilities.
The Lorentzian peak width indicates the lifetime of the eigenstate due to the time-energy uncertainty relation.
The Gaussian peak width indicates the Doppler effect caused by the kinetics of atoms and depends on temperature.
These pieces of information about the electronic states or kinetics of atoms are obtained by identifying peaks from spectra.

It is generally a difficult problem to distinguish each peak from noisy spectra with overlapping peaks.
The simplest solution is least-squares fitting by a gradient method \cite{allen1978deconvolution}.
This type of method has a drawback in that fitting parameters are often trapped at a local minimum or a saddle whenever there is another global minimum in the parameter space.
Moreover, the number of peaks is not always known in practice. 
Bayesian inference, by using a Markov chain Monte Carlo (MCMC) method, provides a superior solution \cite{fischer2001analysis, razul2003bayesian, masson2010dynamics, nagata2012bayesian, mazet2015unsupervised, kasai2016nmr, hong2016automatic, PhysRevC.93.061601, murata2016extraction}.
Although the Bayesian framework enables us to estimate the number of peaks, MCMC methods generally have the limitation of local minima and saddles.
Nagata et al. reported \cite{nagata2012bayesian} that the exchange Monte Carlo method \cite{hukushima1996exchange} (or parallel tempering \cite{geyer1991markov}) can prevent local minima or saddles efficiently and provide a more accurate estimation than the reversible jump MCMC method \cite{green1995reversible} and its extension \cite{jasra2007population}.

We constructed a Bayesian framework for estimating both the noise variance and the number of peaks from spectra with white Gaussian noise by expanding the previous framework by Nagata et al. \cite{nagata2012bayesian}.
The noise variance and the number of peaks are respectively estimated by hyperparameter optimization and model selection.
These estimations are carried out by maximizing a function called the marginal likelihood \cite{efron1973stein, akaike1980likelihood, mackay1992bayesian}, which is a conditional probability of observed data given the noise variance and the number of peaks in our framework.
We provide a straightforward and efficient scheme that calculates this bivariate function by using the exchange Monte Carlo method and the multiple histogram method\cite{ferrenberg1989optimized, kumar1992weighted}.
We also demonstrated our framework through simulation. We show that estimating both the noise variance and the number of peaks prevents overfitting, overpenalizing, and misunderstanding the precision of parameter estimation.

\section{Framework}

\subsection{Models}
An observed spectrum $y \in \mathbb{R}$ is represented by the sum  $f(x; w)$ of single peaks $\phi_k(x; \mu_k,\rho_k)$ and additive noise $\varepsilon$ as
\begin{linenomath*} 
\begin{align}
y &= f(x; w) + \varepsilon, \\
f(x; w) &:= \sum_{k=1}^K a_k \phi_k(x; \mu_k,\rho_k), \\
\phi_k(x; \mu_k,\rho_k) &:= \exp \left[ -\frac{\rho_k}{2} \left(x-\mu_k \right)^2 \right], 
\end{align}
\end{linenomath*} 
where $x \in \mathbb{R}$ denotes energy, frequency, or wave number depending on the case. The parameter set is $w := \{a_k, \mu_k, \rho_k \}_{k=1}^K$, 
where $a_k \geq 0$, $\mu_k \in \mathbb{R}$, and $\rho_k^{-1/2} (\rho_k \geq 0) $ for each $k$ are respectively the intensity, energy level, and peak width. The Gaussian function $\phi_k(x)$ for each $k$ should be replaced with other parametric functions, such as the Lorentzian or Voigt function, depending on the case \cite{tkachenko2006optical, loudon2000quantum}. If the peaks $\phi_k(x)$ are symmetric functions for all $k$ (i.e., their values depend only on the distance from each center), the function $f(x; w)$ is called a radial basis function network
in neural networks and related fields \cite{nagata2012bayesian, broomhead1988radial}. This is the junction of the spectral data analysis and singular learning theory \cite{tokuda2013numerical}.
If the additive noise $\varepsilon$ is assumed to be a zero-mean Gaussian with variance $b^{-1} \geq 0$, the statistical model of the observed spectrum is represented by a conditional probability as
\begin{linenomath*} 
\begin{align}
p(y \mid x,w,b) := \sqrt{\frac{b}{2\pi}} \exp \left \{ - \frac{b}{2} [y - f(x; w)]^2 \right \}, \label{eq:stat_model}
\end{align} 
\end{linenomath*} 
where $y$ is taken as a random variable. This Gaussian distribution $p(y \mid x, w,b)$ is valid if the thermal noise is dominant.
The parameter set $w$ is also regarded as a random variable from the Bayesian viewpoint. 
The probability density function of $w$, called the {\it prior} density, is heuristically modeled as
\begin{linenomath*} 
\begin{align}
\varphi(w \mid K) &:= \prod_{k=1}^{K} \varphi \left(a_k \right) \varphi \left(\mu_k \right) \varphi \left(\rho_k \right), \\
\varphi \left(a_k \right) &:= \kappa \exp(-\kappa a_k), \label{eq: sparse1} \\
\varphi \left(\mu_k \right) &:= \sqrt{\frac{\alpha}{2 \pi}} \exp \left[ -\frac{\alpha}{2} (\mu_k - \mu_0)^2 \right] \label{eq: flat} \\
\varphi \left(\rho_k \right) &:= \nu \exp \left(- \nu \rho_k \right), \label{eq: sparse2}
\end{align}
\end{linenomath*} 
where $\kappa>0$, $\mu_0 \in \mathbb{R}$, $\alpha>0$, and $\nu>0$ are hyperparameters. 
This prior density modeling is a special case of that by Nagata et al. \cite{nagata2012bayesian}.
Equation (\ref{eq: sparse1}) promotes the sparsity of $a_k$.
Equation (\ref{eq: flat}) is regarded as an almost flat prior density if $\alpha$ is sufficiently small. 
These prior density models can be replaced with any other model without loss of generality in our framework.

\subsection{Bayesian formalization}
The conditional probability density function of $w$ given samples $D:=\{X_i, Y_i\}_{i=1}^n$, set as $X_1 < X_2 < \cdots < X_n$ for the sake of convenience, is represented by Bayes' theorem as
\begin{linenomath*} 
\begin{align}
p(w \mid D, K, b) &= \frac{1}{Z_n(K,b)} \prod_{i=1}^n p(Y_i \mid X_i ,w,b) \varphi(w \mid K) \\
&= \frac{1}{\tilde{Z}_n(K, b)} \exp \left[ -nb E_n(w) \right] \varphi(w \mid K), \\ 
Z_n(K,b) &:= \int dw \prod_{i=1}^n p(Y_i \mid X_i ,w,b) \varphi(w \mid K) \label{eq:marginal} \\
&= \left( \frac{b}{2\pi} \right)^{\frac{n}{2}} \tilde{Z}_n(K,b), \\
\tilde{Z}_n(K,b) &:= \int dw \exp \left[ -nb E_n(w) \right] \varphi(w \mid K), \\
E_n(w) &:= \frac{1}{2n} \sum_{i=1}^n \left[ Y_i -f(X_i; w) \right]^2, \label{eq:error}
\end{align}
\end{linenomath*} 
where the functions $p(w \mid D, K, b)$ and $Z_n(K,b)$ are respectively called the {\it posterior} density and marginal likelihood. Note that the function $Z_n(K,b) = p(\{Y_i\}_{i=1}^n \mid \{X_i\}_{i=1}^n K,b)$ is a probability density but $\tilde{Z}_n(K,b)$ is not.
Bayes free energy $F_n(K,b)$ is defined as
\begin{linenomath*} 
\begin{align}
F_n(K,b) &:= -\log Z_n(K,b) \\
&= b \tilde{F_n}(K,b) -\frac{n}{2} (\log b - \log 2 \pi), \\
\tilde{F_n}(K,b) &:= -\frac{1}{b} \log \tilde{Z}_n(K,b). 
\end{align}
\end{linenomath*} 
Note that Nagata et al. regarded $b \tilde{F_n}(K,b)$ as Bayes free energy for the sake of convenience \cite{nagata2012bayesian} since the noise variance is treated as a known constant.
We also assume the case in which there are no peaks as $K=0$ (see Appendix A).
In terms of the empirical Bayes (or type II maximum likelihood) approach \cite{efron1973stein, akaike1980likelihood, mackay1992bayesian}, empirical Bayes estimators of $K$ and $b$ are given by
\begin{linenomath*} 
\begin{align}
(\hat{K},\hat{b}) &:= \argmax_{K, b} Z_n(K,b) \\
&=\argmin_{K, b} F_n(K,b).
\end{align}
\end{linenomath*} 
The hierarchical Bayes approach \cite{gelman2013bayesian} is also tractable in our framework (see Appendix B).
The partial derivative of $F_n(K,b)$ with respect to the variable $b$ is obtained as
\begin{linenomath*} 
\begin{align}
\frac{\partial F_n}{\partial b} &= n \left[ \langle E_n(w) \rangle_b -\frac{1}{2b} \right], \label{eq: derivative}
\end{align}
\end{linenomath*} 
where $\langle Q \rangle_{b}$ denotes the posterior mean of an arbitrary quantity $Q \in \mathbb{R}$ over $p(w \mid D, K, b)$. 
If $b=\hat{b}$ is a stationary point of $F_n(K,b)$, then the following equation is satisfied:
\begin{linenomath*} 
\begin{align}
\langle E_n(w) \rangle_{\hat{b}} = \frac{1}{2\hat{b}}. \label{eq:deri2}
\end{align}
\end{linenomath*} 
The Bayes estimator of $w$ is given by $\hat{w} := \left \{ \langle a_k \rangle_{\hat{b}}, \langle \mu_k \rangle_{\hat{b}}, \langle \rho_k \rangle_{\hat{b}} \right \}_{k=1}^{\hat{K}}$ with the standard deviation $\sqrt{\langle {Q'}^2 \rangle_{\hat{b}} - {\langle Q' \rangle_{\hat{b}}}^2}$ for each parameter $Q' \in w$ if $\hat{K}>0$.
However, $(\hat{K}, \hat{b})$ cannot be derived in this case since $F_n(K,b)$ and $\langle E_n(w) \rangle_{b}$ are analytically intractable for our model.

\subsection{Exchange Monte Carlo method}
In practice, we calculate $F_n(K,b)$ and $\langle E_n(w) \rangle_{b}$ by using the exchange Monte Carlo method, which efficiently enables sampling from $p(w \mid D, K, b)$ at $b \in \{b_l\}_{l=1}^L$ without knowing $Z_n(K,b)$ or $F_n(K,b)$. The target density is a joint probability density as
\begin{linenomath*} 
\begin{align}
p \left( \{w_l\}_{l=1}^L \mid D, K, \{b_l\}_{l=1}^L \right) &:= \prod_{l=1}^L p(w_l \mid D, K,b_l),
\end{align}
\end{linenomath*} 
where $w_l$ is the parameter set at $b_l$. Each density $p(w_l \mid D, K,b_l)$ is called a {\it replica}. Sequence $\{b_l\}_{l=1}^L$ is set as $0 = b_1 < b_2 < \cdots < b_L$ for the sake of convenience. Note that the variable $b$ is replaced with the inverse temperature $\beta$ of Nagata et al.'s formulation\cite{nagata2012bayesian}.
The variable $b$ works as quasi-inverse temperature and varies the substantial support of the posterior density $p(w \mid D, K, b)$. The state exchange between high- and low-temperature replicas enables the escape from local minima or saddles in the parameter space. The sampling procedure includes the two following steps.
\begin{itemize}
\item State update in each replica \\
Simultaneously and independently update state $w_l$ subject to $p(w_l \mid D, K,b_l)$ using the Metropolis algorithm \cite{metropolis1953equation}.
\item State exchange between neighboring replicas \\
Exchange states $w_l$ and $w_{l+1}$ at every step subject to the probability $u(w_{l+1}, w_{l}, b_{l+1}, b_{l})$ as
\begin{linenomath*} 
\begin{align}
u(w_{l+1}, w_{l}, b_{l+1}, b_{l}) &:= \min \left[1, v(w_{l+1}, w_{l}, b_{l+1}, b_{l}) \right], \label{eq:balance} \\
v(w_{l+1}, w_{l}, b_{l+1}, b_{l}) &:= \frac{p(w_{l+1} \mid D,K,b_{l}) p(w_{l} \mid D,K,b_{l+1})}{p(w_{l} \mid D,K,b_{l}) p(w_{l+1} \mid D,K,b_{l+1})} \\
&= \exp \left \{ n (b_{l+1} - b_{l}) [E_n(w_{l+1}) - E_n(w_{l})] \right \},
\end{align}
\end{linenomath*} 
where Eq. (\ref{eq:balance}) ensures a detailed balance condition.
\end{itemize}
A straightforward way of computing $\tilde{F}_n(K,b_l)$ via the exchange Monte Carlo method is bridge sampling \cite{meng1996simulating, gelman1998simulating}, in which $\tilde{F}_n(K,b_l)$ is expressed as
\begin{linenomath*} 
\begin{align}
\tilde{F}_n(K,b_l) &= - \frac{1}{b_l} \log \prod_{l'=1}^{l-1} \frac{\tilde{Z}(K,b_{l'+1})}{\tilde{Z}(K,b_{l'})} \\
&= - \frac{1}{b_l} \sum_{l'=1}^{l-1} \log \langle \exp [-n (b_{l'+1} - b_{l'}) E_n(w_{l'}) ] \rangle_{b_{l'}}, \label{eq:bridge}
\end{align}
\end{linenomath*} 
where $\langle Q_l \rangle_{b_l}$ for the arbitrary quantity $Q_l \in \mathbb{R}$ at the $l^{\rm th}$ replica is approximated by the mean of an MCMC sample $\{Q_{l, m}\}_{m=1}^{M_{l}}$ as
\begin{linenomath*} 
\begin{align}
\langle Q_l \rangle_{b_l} &= \frac{1}{M_{l}}\sum_{m=1}^{M_{l}} Q_{l, m}. \label{eq:thermal_ave}
\end{align}
\end{linenomath*} 
However, $\hat{b}$ is not easy to accurately calculate using only the above scheme since $\{b_l\}_{l=1}^L$ is a discrete set, whereas $b$ is a continuous variable.

\subsection{Multiple histogram method}
We interpolate $\{F_n(K,b_l)\}_{l=1}^L$ or $\{\langle E_n(w) \rangle_{b_l}\}_{l=1}^L$ with respect to $b=b' \in (b_l,b_{l+1})$ for any $l$ via the multiple histogram method. The density of states is defined and estimated by
\begin{linenomath*} 
\begin{align}
g(E;K) &:= \int dw \delta [E-E_n(w)] \varphi (w \mid K) \\
&= \frac{\sum_{l=1}^L N_{l}(E)}{\sum_{l'=1}^L M_{l'} \tilde{Z}_n(K,b_{l'})^{-1} \exp(-n b_{l'} E)},
\end{align}
\end{linenomath*} 
then we obtain 
\begin{linenomath*} 
\begin{align}
\tilde{Z}_n(K,b) &= \int dE g(E;K) \exp (-nbE) \\
&= \sum_{l=1}^L \sum_{m=1}^{M_{l}} \frac{1}{\sum_{l'=1}^L M_{l'} \tilde{Z}_n(K,b_{l'})^{-1} \exp \left[ n (b-b_{l'}) E_{l, m} \right]}, \label{eq:MHR}
\end{align}
\end{linenomath*} 
where $N_l(E)dE$ and $E_{l,m}$ are respectively the histogram of $E \geq 0$ at the $l^{\rm th}$ replica and the value of $E$ at the $m^{\rm th}$ snapshot of the $l^{\rm th}$ replica in an MCMC simulation, i.e., $\int dE N_l(E) = M_l$. The values of $\{\tilde{Z}_n(K,b_l)\}_{l=1}^L$ are determined self-consistently by iterating Eq. (\ref{eq:MHR}) with $b=b_l$. We take $\exp[-b_l \tilde{F}_n(K,b_l)]$ computed via Eq. (\ref{eq:bridge}) as the initial values for the sake of convenience. Given $\{\tilde{Z}_n(K,b_l)\}_{l=1}^L$, we then calculate $\tilde{Z}_n(K,b)$ as $b=b'$ via Eq. (\ref{eq:MHR}) again. The above procedure can be appropriately generalized to treat multidimensional histograms such as $N_l(E,Q)dEdQ$ \cite{newman1999monte}. Then, the posterior mean of an arbitrary quantity is calculated as  
\begin{linenomath*} 
\begin{align}
\langle Q \rangle_b &= \frac{1}{\tilde{Z}_n(K,b)} \sum_{l=1}^L \sum_{m=1}^{M_{l}} \frac{Q_{l, m}}{\sum_{l'=1}^L M_{l'} \tilde{Z}_n(K,b_{l'})^{-1} \exp \left[ n (b-b_{l'}) E_{l, m} \right]}, \label{eq:MHR_ave}
\end{align}
\end{linenomath*} 
where $Q_{l,m}$ is the value of $Q$ at the $m^{\rm th}$ snapshot of the $l^{\rm th}$ replica in an MCMC simulation.
We calculate $\langle E_n(w) \rangle_b$ via Eq. (\ref{eq:MHR_ave}) and solve Eq. (\ref{eq:deri2}) numerically by the bisection method.
Then, $\hat{w}$ with the standard deviation of each parameter is also calculated via Eq. (\ref{eq:MHR_ave}).
The posterior density of arbitrary quantities can also be interpolated with respect to $b=b'$ in the same way (see Appendix C).

\begin{figure}[t]
\begin{tabular}{c}
\includegraphics[width=7cm]{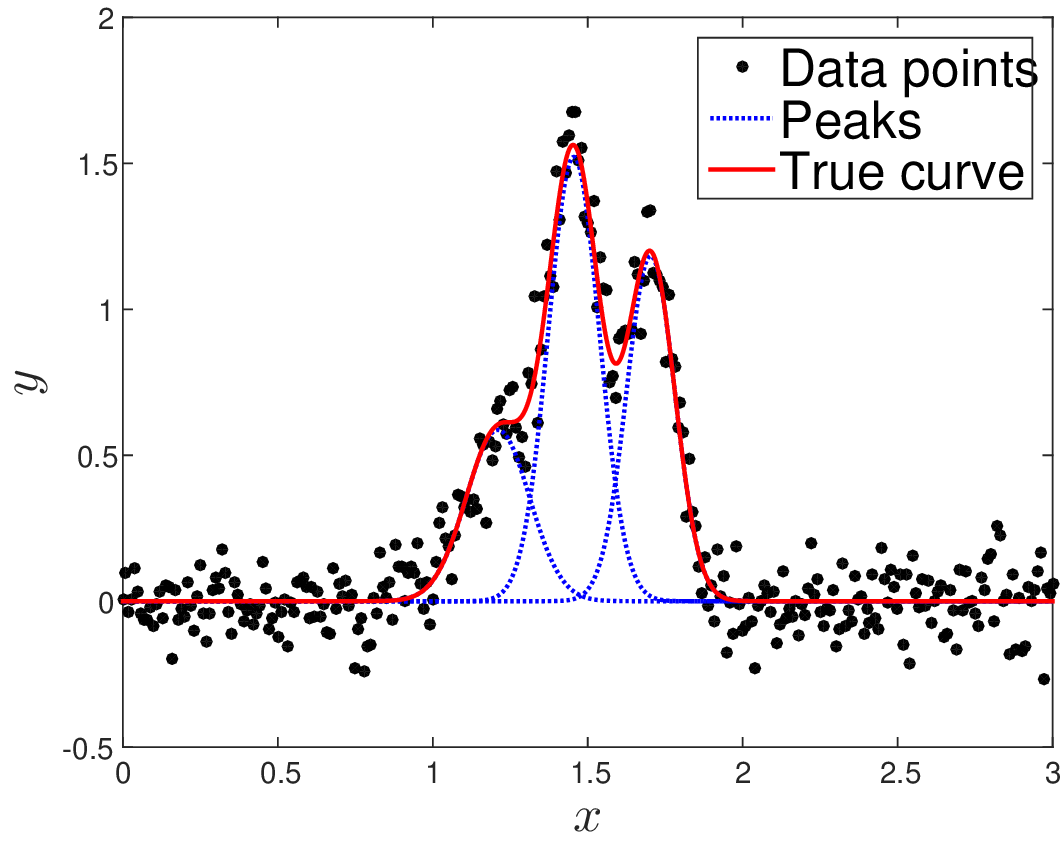}
\end{tabular}
\caption{(Color online) Synthetic data. The horizontal and vertical axes respectively 
represent the input $x$ and output $y$. The black dots show synthetic data $D=\{X_i, Y_i\}_{i=1}^n$. The red solid line and blue dotted ones respectively show the true curve $y=f(x; w_0)$ and the Gaussian peaks $y=\phi_k(x; {\mu_k}^*, {\rho_k}^*)$.} 
\label{fig:data}
\end{figure}

\section{Demonstration}
\label{sec:demo}
We demonstrated how efficient our framework is through simulation in which the same synthetic data as used by Nagata et al. \cite{nagata2012bayesian} were used.
The synthetic data $D=\{X_i, Y_i\}_{i=1}^n$ shown in Fig. \ref{fig:data} were generated from the true probability density as
\begin{linenomath*} 
\begin{align}
q(y \mid x, w_0,b_0) := \sqrt{\frac{b_0}{2\pi}} \exp \left \{ - \frac{b_0}{2} [y - f(x; w_0)]^2 \right \}, \label{eq:true}
\end{align} 
\end{linenomath*} 
where $b_0>0$ and $w_0 := \{{a_k}^*, {\mu_k}^*, {\rho_k}^* \}_{k=1}^{K_0}$ are respectively the true inverse noise variance and true parameter set, as in Tables \ref{tb:hyperparameters} and \ref{tb:parameters}.
The inputs $\{X_i\}_{i=1}^n$ were linearly spaced in the interval $[X_1, X_n]=[0, 3]$ with  spectral resolution $\Delta x = 0.01$, where the number of samples was $n=301$.
The sequence $\{b_l\}_{l=2}^L$ were logarithmically spaced in the interval $[nb_2, nb_L]=[10^{-4}, 10^8]$, where the number of replicas was $L=400$.
The model size $K$ was set as integers from $0$ to $5$.
The hyperparameters were $\kappa=1.7$, $\mu_0=1.5$, $\alpha=0.4$, and $\nu=0.01$ in the  heuristics. The total number of MCMC sweeps was 100,000 including 50,000 burn-in sweeps: an MCMC sample $\{w_{l,m}\}_{m=1}^{M_l}$ of size $M_l=50,000$ for every $b_l$ was obtained. 
The estimators are listed in Tables \ref{tb:hyperparameters} and \ref{tb:parameters}, where $\rho_k$ was converted into an inverse square-root scale for comparison. Every true value of the parameter lies within two standard deviations. 

\begin{table}[t]
\caption{Number of peaks and inverse noise variance.}
\begin{tabular}{ccc}
\hline
 & $K$ & $b$ \\
\hline
Estimated & $3$ & $1.029406\times10^2$ \\
True & $3$ & $1.00000\times10^2$ \\
\hline
\end{tabular}
\label{tb:hyperparameters}
\end{table}

\begin{table}[t]
\caption{Parameters of each Gaussian peak.}
\begin{tabular}{c|cccc}
\hline
\multicolumn{2}{c}{ } & $a_k$ & $\mu_k$ & ${\rho_k}^{-1/2}$ \\
\hline
Mode 1 & Estimated & $0.5794 \pm 0.0542$ & $1.2571 \pm 0.0395$ & $0.144132 \pm 0.025711$ \\
$(k=1)$ & True & $0.587$ & $1.210$ & $0.10223$ \\
\hline
Mode 2 & Estimated & $1.3514 \pm 0.1518$ & $1.4605 \pm 0.0043$ & $0.07606120 \pm 0.00604382$ \\
$(k=2)$ & True & $1.522$ & $1.455$ & $0.0825244$ \\
\hline
Mode 3 & Estimated & $1.1600 \pm0.0483$ & $1.7032 \pm 0.0044$ & $0.08175039 \pm 0.00407585$ \\
$(k=3)$ & True & $1.183$ & $1.703$ & $0.0779755$ \\
\hline
\end{tabular}
\label{tb:parameters}
\end{table}

\begin{figure}[t]
\begin{tabular}{ll}
(A-1) & (A-2) \\
\vspace{-8pt}
\\
\includegraphics[width=7cm]{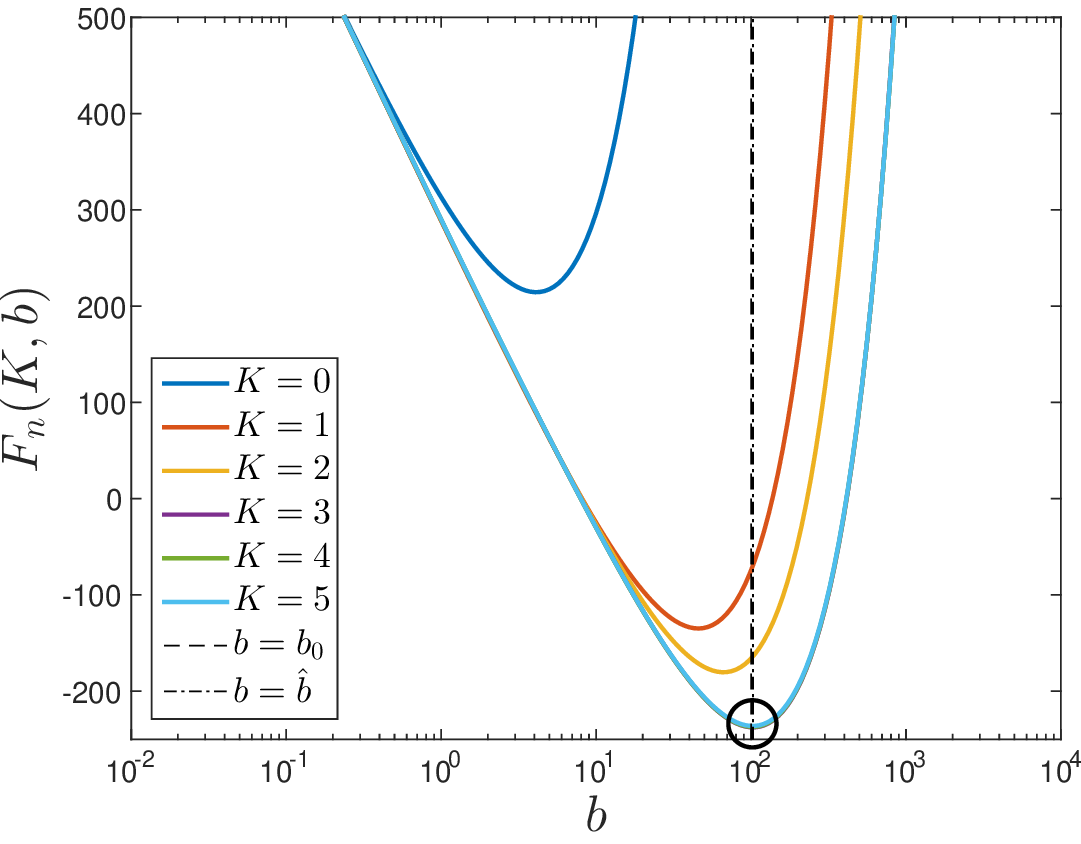}
&
\includegraphics[width=7cm]{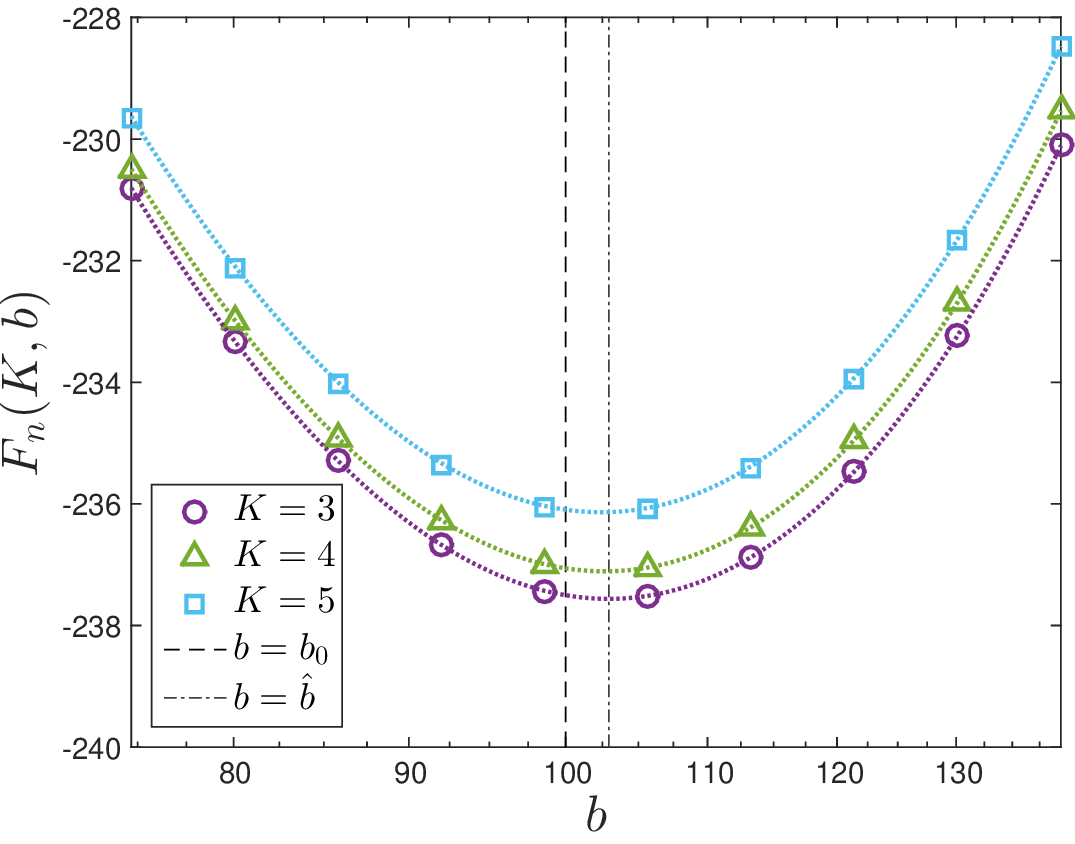}
\\
(B-1) & (B-2) \\
\vspace{-8pt}
\\
\includegraphics[width=7cm]{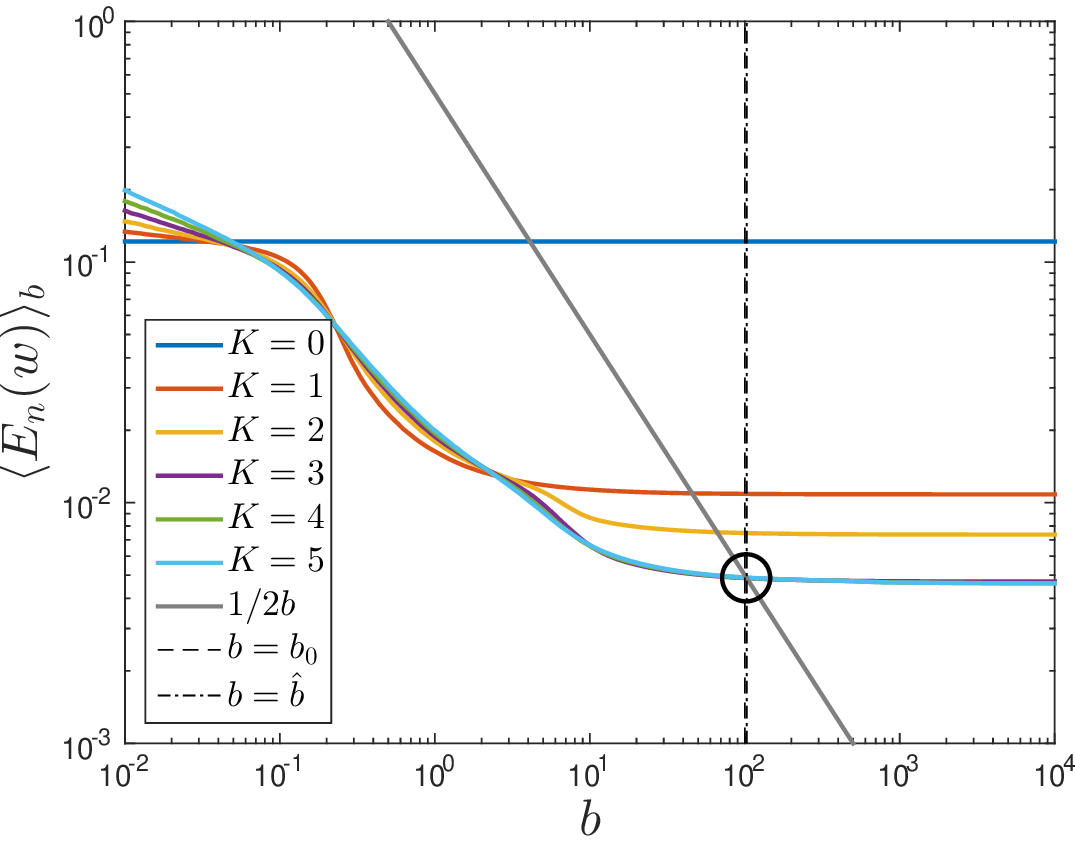}
&
\includegraphics[width=7cm]{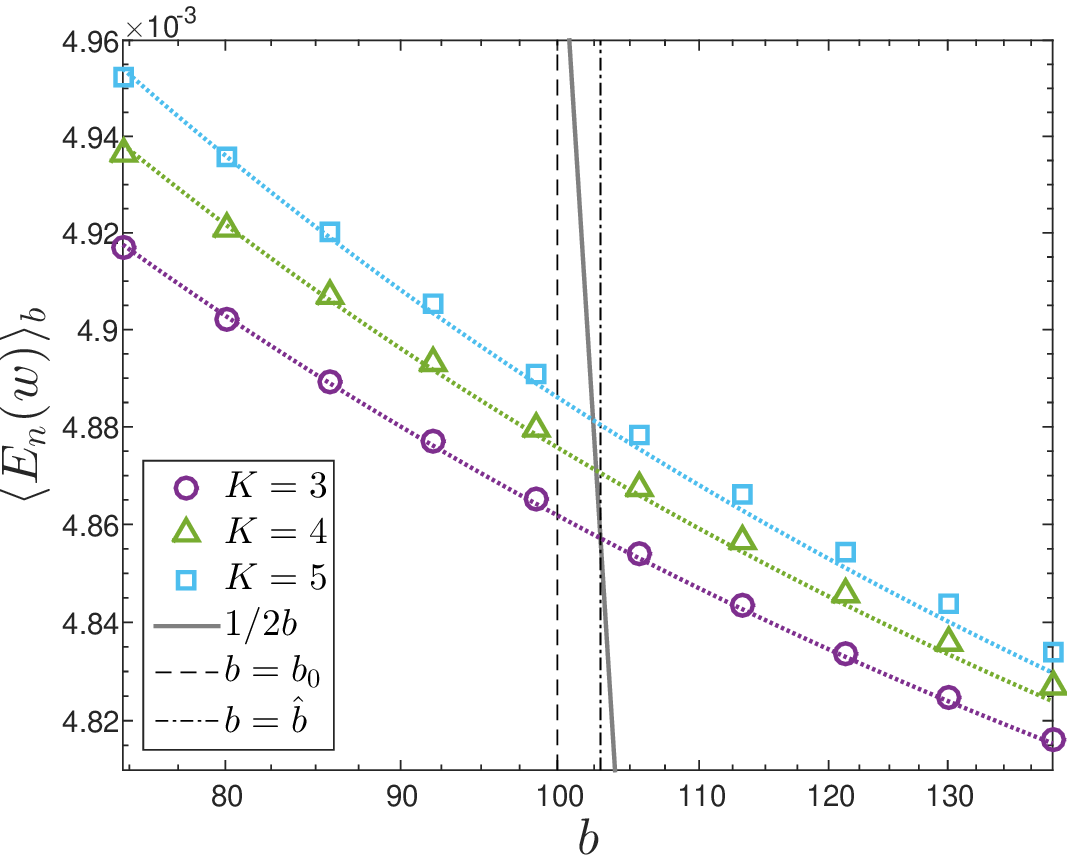}
\end{tabular}
\caption{(Color online) (A) Bayes free energy and (B) posterior mean of mean square error. The horizontal axes represent $b$ on a log scale. The colored solid lines show $F_n(K,b_l)$ for each $K$ in (A) and $\langle E_n(w) \rangle_{b_l}$ for each $K$ on a log scale in (B). The three lines of $K \geq 3$ almost overlap in (A-1) and (B-1) whose enlarged views around black circles are respectively shown in (A-2) and (B-2). The colored markers in (A-2) and (B-2) respectively indicate $F_n(K,b_l)$ as in (A-1) and $\langle E_n(w) \rangle_{b_l}$ as in (B-1). The colored dotted lines in (A-2) and (B-2) indicate the interpolated values. The gray solid lines in (B) show the function $1/2b$. The vertical black dashed lines and vertical black dash-dotted ones respectively show the true value $b=b_0$ and the estimated value $b=\hat{b}$.}
\label{fig:state_function}
\end{figure}

First, we discuss how to estimate both the noise variance and the number of peaks.
(A) Bayes free energy and (B) the posterior mean of the mean square error are shown in Fig. \ref{fig:state_function}. The horizontal axes represent $b$ on a log scale. The colored solid lines show $F_n(K,b_l)$ calculated via Eq. (\ref{eq:bridge}) for each $K$ in (A) and $\langle E_n(w) \rangle_{b_l}$ calculated via Eq. (\ref{eq:thermal_ave}) for each $K$ on a log scale in (B). The three lines of $K \geq 3$ almost overlap in (A-1) and (B-1), whose enlarged views around the black circles are respectively shown in (A-2) and (B-2). The colored markers in (A-2) and (B-2) respectively indicate $F_n(K,b_l)$ as in (A-1) and $\langle E_n(w) \rangle_{b_l}$ as in (B-1).
The colored dotted lines in (A-2) and (B-2) respectively indicate the interpolated values calculated via Eqs. (\ref{eq:MHR}) and (\ref{eq:MHR_ave}).
The gray solid lines in (B) show the function $1/2b$. The vertical black dashed lines and vertical black dash-dotted ones respectively show the true value $b=b_0$ and the estimated value $b=\hat{b}$.
There is a minimum point of $F_n(K,b)$ depending on each value of $K$, i.e., the probability density $p(K,b \mid D)$ has a maximum at this point (see Appendix B).
In this case, Eq. (\ref{eq:deri2}) holds at the intersection of the purple dotted line and the gray solid line shown in (B-2).

\begin{figure}[t]
\begin{tabular}{c}
\includegraphics[width=7cm]{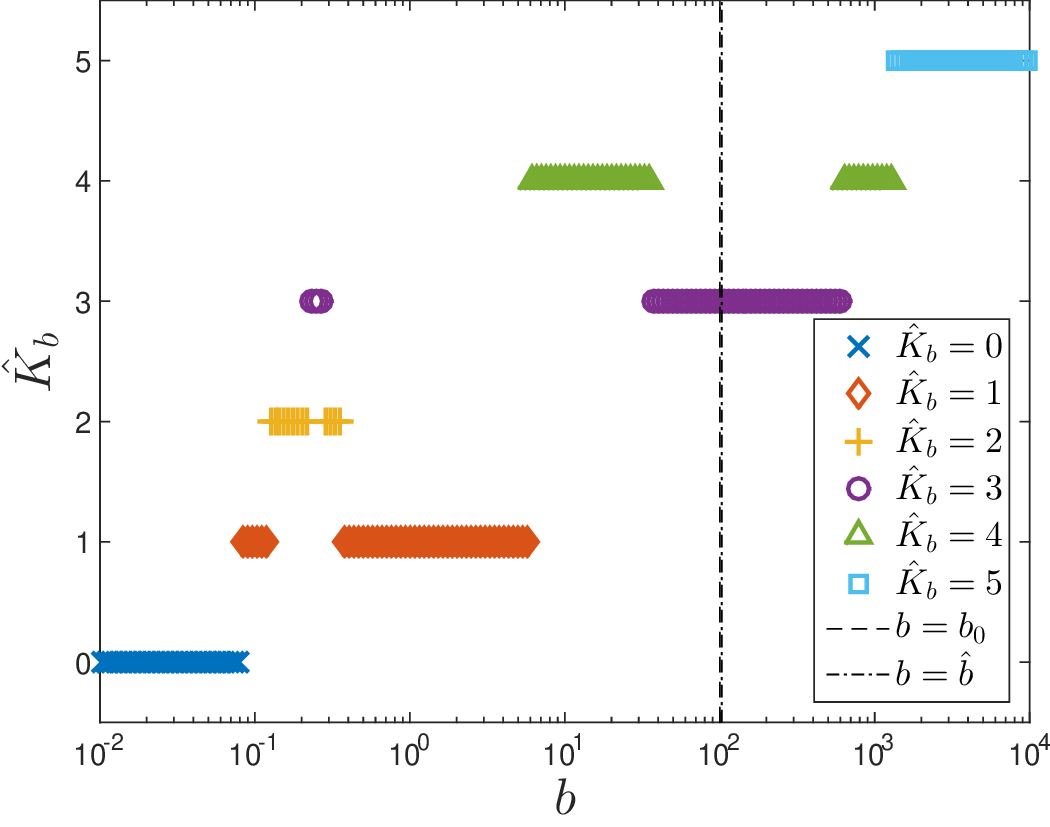}
\end{tabular}
\caption{(Color online) Dependence of model selection on $b$. The horizontal axis represents $b$ on a log scale. The estimated model size $\hat{K}_b$ that minimizes $F_n(K, b)$ for each $b$ is plotted as colored marker. The vertical black dashed line and the vertical black dash-dotted one respectively show the true value $b=b_0$ and the estimated value $b=\hat{b}$.}
\label{fig:single}
\end{figure}

Second, we discuss the validity of our framework. The dependence on $b$ in the model selection is shown in Fig. \ref{fig:single}. The horizontal axis represents $b$ on a log scale. The colored markers show the estimated model size $\hat{K}_b$ that minimizes $F_n(K, b_l)$ for each $b_l$ as 
\begin{linenomath*} 
\begin{align}
\hat{K}_b &:= \argmin_{K} F_n(K, b_l) \\
&= \argmin_{K} \tilde{F}_n(K, b_l).
\end{align}
\end{linenomath*} 
Note that $\hat{K}_{b_0} = {\rm arg~min}_{K} \tilde{F}_n(K, b_0)$ is regarded as the optimal number of peaks in Nagata et al.'s framework \cite{nagata2012bayesian}.
The vertical black dashed line and the vertical black dash-dotted one respectively show the true value $b=b_0$ and the estimated value $b=\hat{b}$.
Although $\hat{K}_b$ for each value of $b$ depends on the noise realization, as Nagata et al. showed in the case of $b=b_0$ \cite{nagata2012bayesian}, $\hat{K}_b$ also changes depending on the value of $b$.
There is a rough trend, explained by the asymptotic form of $\tilde{F}_n(K, b)$, in which $\hat{K}_b$ becomes larger as $b$ increases.
If the sample size $n$ is sufficiently large, $\tilde{F}_n(K, b)$ is expressed as
\begin{linenomath*} 
\begin{align}
\tilde{F}_n(K, b) &= nE_n(w_0) + \frac{\lambda}{b} \log nb + \frac{1}{b} O_p(\log \log nb), \label{eq:Helmholtz}
\end{align}
\end{linenomath*} 
where $w_0$ is the parameter set that minimizes the Kullback--Leibler divergence of a statistical model from a true distribution, and $\lambda>0$ is a rational number called the real log canonical threshold (RLCT) \cite{watanabe2001algebraic, watanabe2009algebraic}.
The RLCT is determined by the pair of a statistical model and true distribution, and the ones determined by Eqs. (\ref{eq:stat_model}) and (\ref{eq:true}) are clarified for several cases of $(K, K_0)$ with $b=b_0$ \cite{tokuda2013numerical}. 
The values $E_n(w_0)$ and $\lambda$ respectively become larger and smaller as $K$ increases.
The term $nE_n(w_0)$ dominantly works for model selection for large $b$: overfitting occurs.
The term $\lambda \log nb$ dominantly works for small $b$: overpenalizing occurs.
A moderate model is estimated under the moderate value of $b$.
Estimating the optimal value of $b$ is indispensable, and this result shows the validity of our framework.

\begin{figure}[t]
\begin{tabular}{llll}
\multicolumn{2}{l}{(A)} & \multicolumn{2}{l}{(B)} \\
\vspace{-8pt}
\\
\multicolumn{2}{l}{\includegraphics[height=5.5cm]{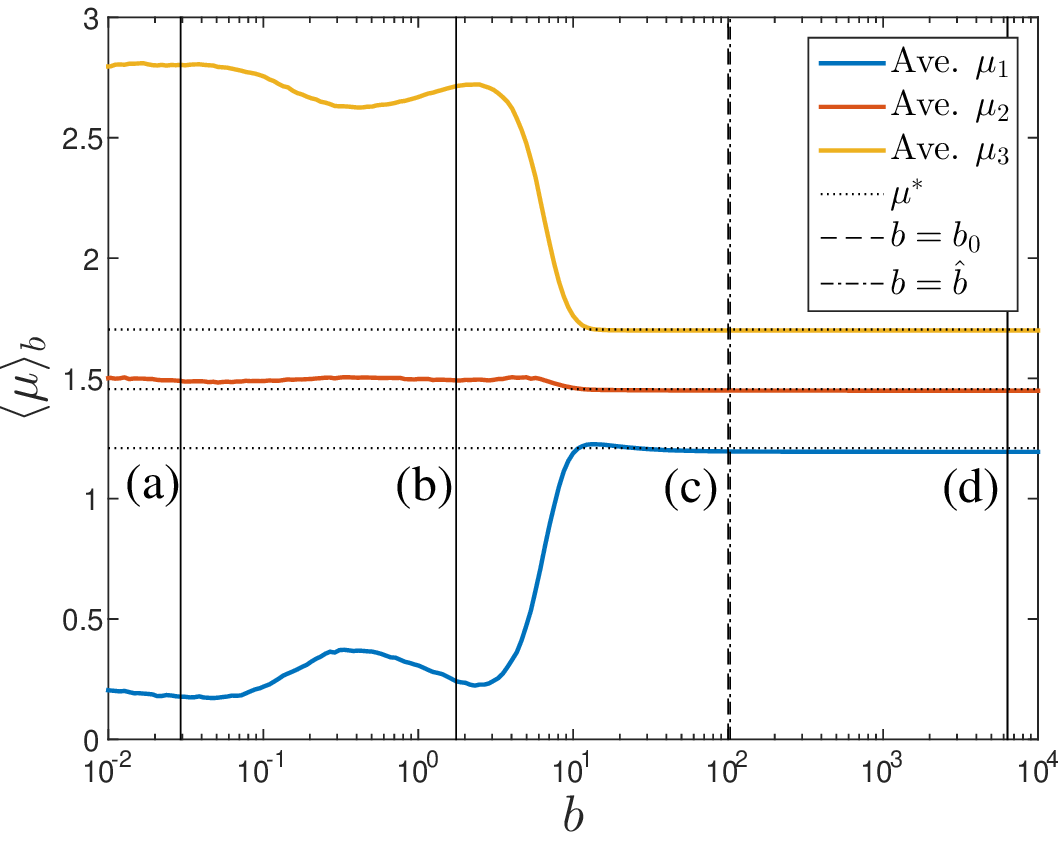}}
&
\multicolumn{2}{l}{\includegraphics[height=5.5cm]{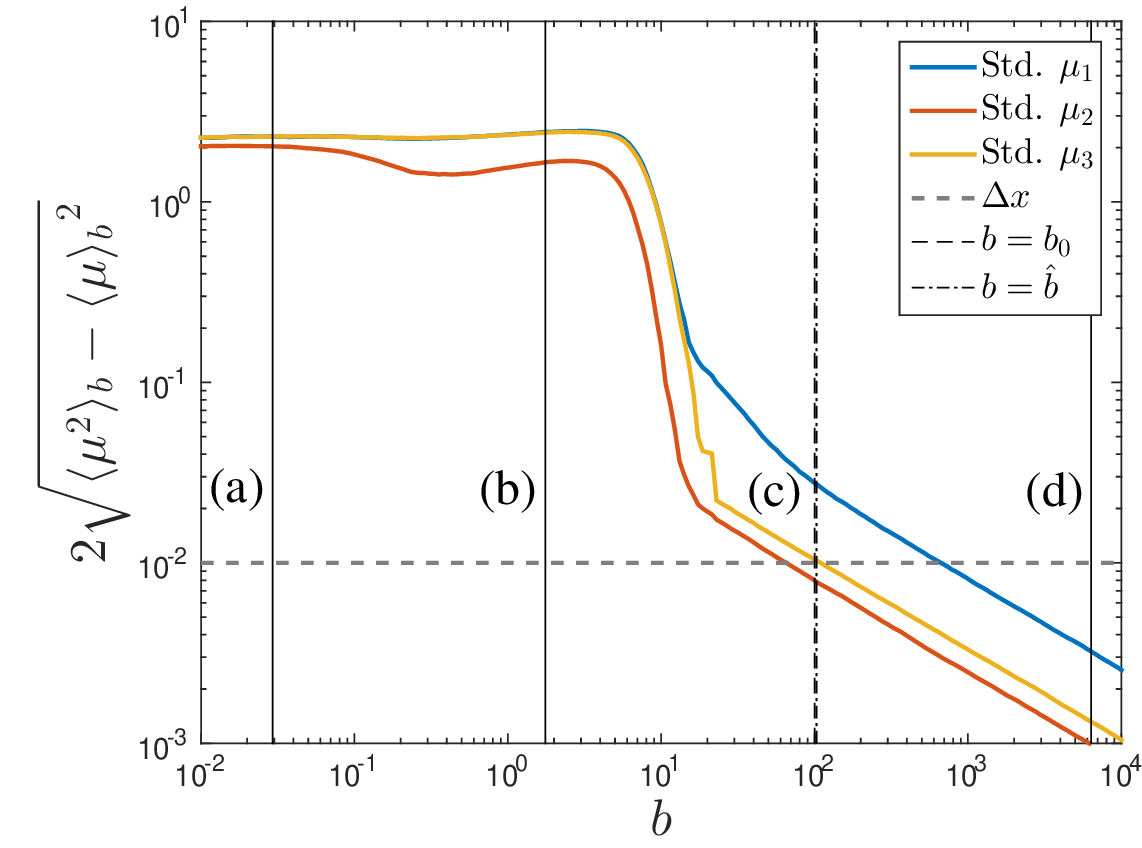}}
\\
(a) $b=2.925210\times10^{-2}$ & (b) $b=1.758132\times10^0$ & (c) $b=\hat{b}$ & (d) $b=6.350977\times10^3$ \\
\vspace{-10pt}
\\
\includegraphics[height=2.75cm]{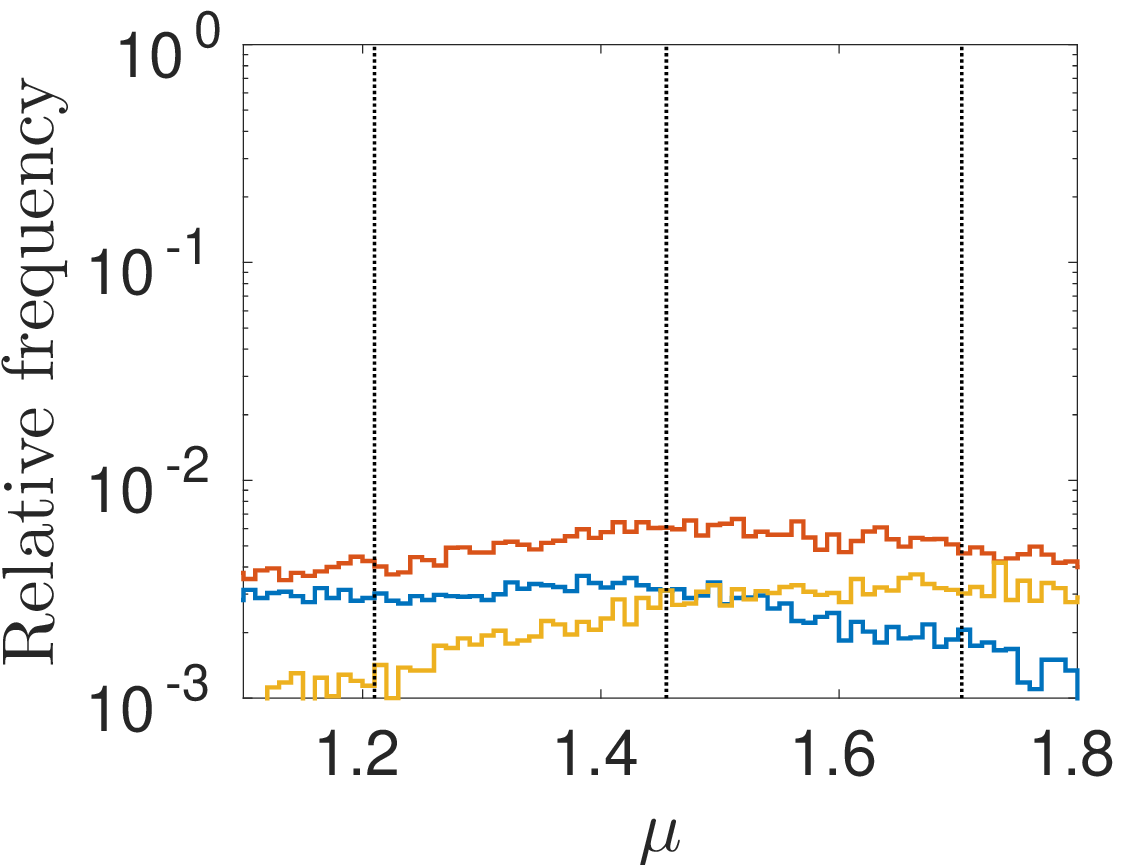}
&
\includegraphics[height=2.75cm]{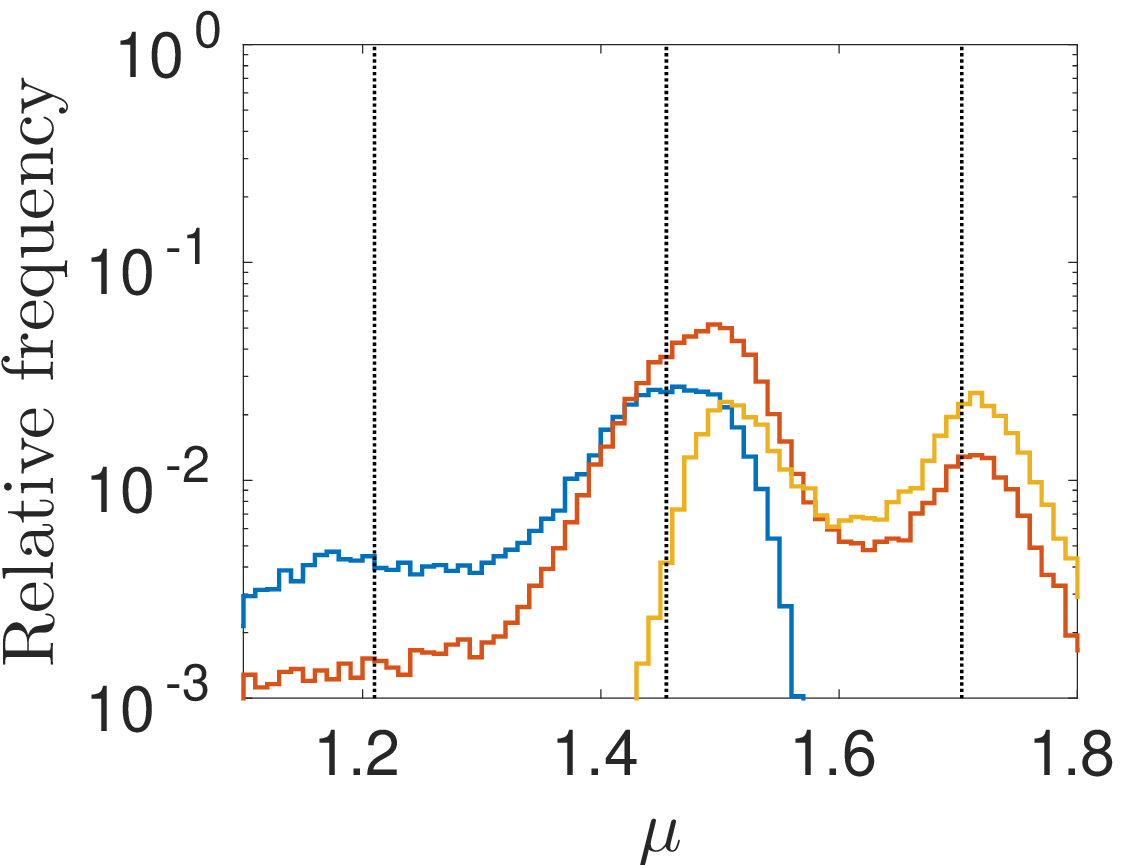}
&
\includegraphics[height=2.75cm]{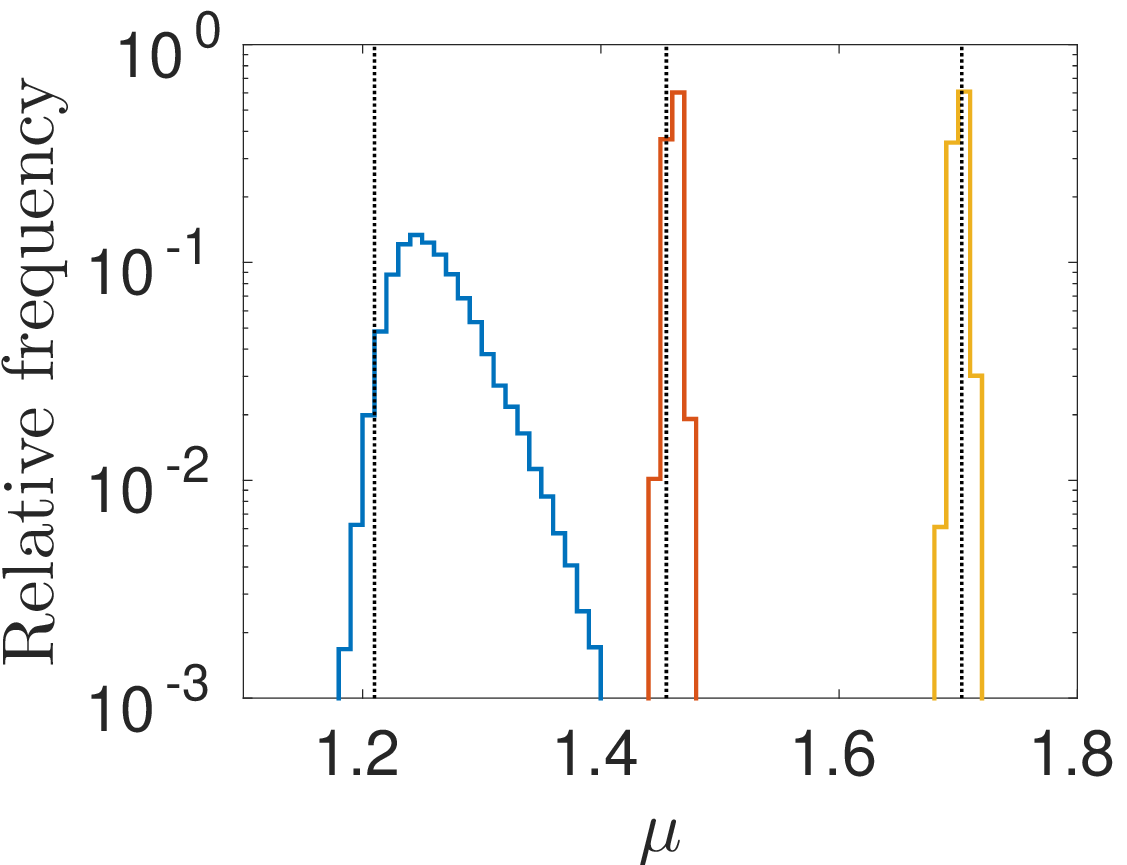}
&
\includegraphics[height=2.75cm]{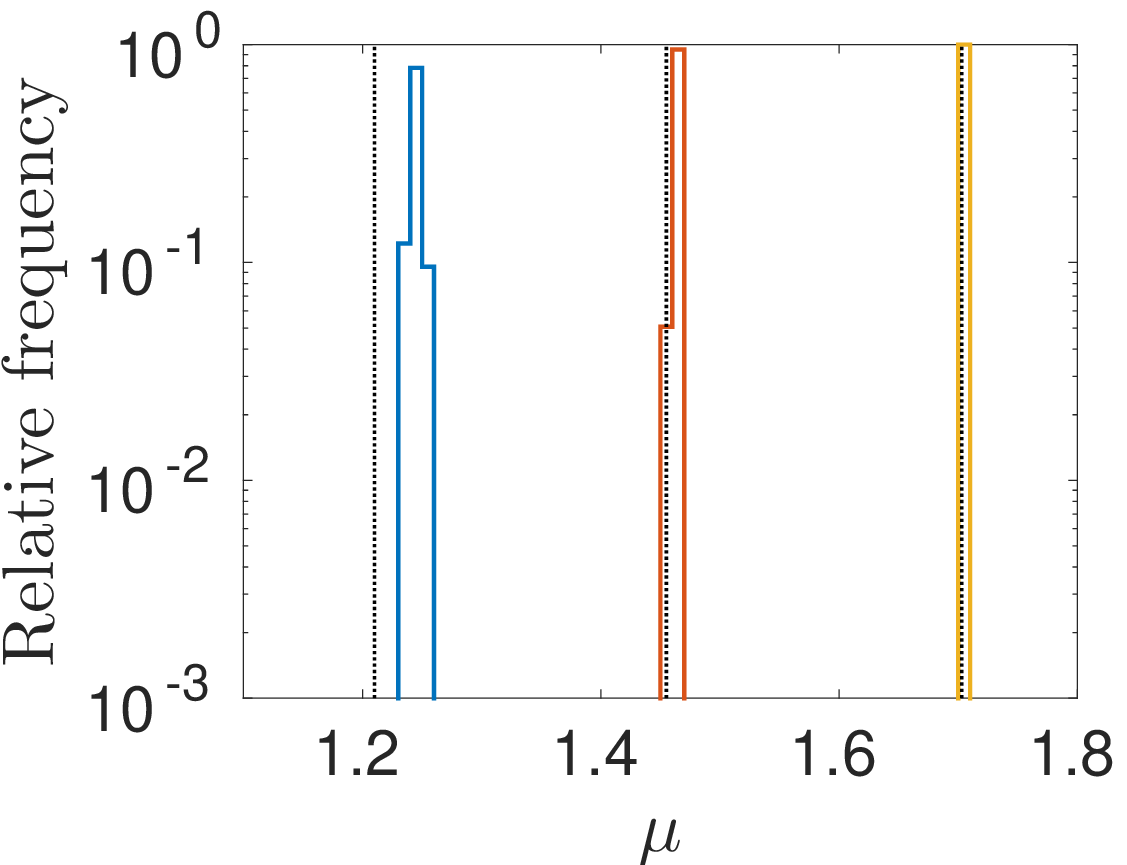}
\end{tabular}
\caption{(Color online) (A) Posterior mean of $\mu_k$, (B) posterior standard deviation of $\mu_k$, and (a-d) marginal posterior distribution of $\mu_k$ when $K=K_0=3$. The horizontal axes in (A-B) represent $b$ on a log scale. The colored solid lines show $\langle \mu_k \rangle_{b_l}$ for each $k$ in (A) and $2 \sqrt{\langle {\mu_k}^2 \rangle_{b_l} - {\langle \mu_k \rangle_{b_l}}^2}$ for each $k$ on a log scale in (B). The vertical black dashed lines and the vertical black dash-dotted ones respectively show the true value $b=b_0$ and the estimated value $b=\hat{b}$. The horizontal black dotted lines in (A) show the true value ${\mu_k}^*$ for each $k$ and the horizontal gray dashed line in (B) shows $\Delta x$. The vertical black solid lines in (A-B) correspond to each value of $b$ in (a-d). The histograms (a-d) of $\mu_k$ show the marginal posterior distribution of $\mu_k$ for each $b$, where the coloring for each $\mu_k$ follows that in (A-B). The horizontal axes in (a-d) represent $\mu_k$, and the vertical ones represent relative frequency on a log scale. The vertical black dotted lines also show the true value ${\mu_k}^*$ for each $k$, as in (A).} 
\label{fig:posterior}
\end{figure}

Finally, we discuss the validity of our framework from another viewpoint.
(A) The posterior mean of $\mu_k$, (B) the posterior standard deviation of $\mu_k$, and (a-d) the marginal posterior distribution of $\mu_k$ when $K=K_0=3$ are shown in Fig. \ref{fig:posterior}. The horizontal axes in (A-B) represent $b$ on a log scale. The colored solid lines show $\langle \mu_k \rangle_{b_l}$ for each $k$ in (A) and $2 \sqrt{\langle {\mu_k}^2 \rangle_{b_l} - {\langle \mu_k \rangle_{b_l}}^2}$ for each $k$ in log scale in (B). These values were calculated via Eq. (\ref{eq:thermal_ave}). The identification of mode $k$ was reassigned by sorting the MCMC sample $\{\mu_{k,l,m}\}_{k=1}^{3}$ into $\mu_{1,l,m}< \mu_{2,l,m}< \mu_{3,l,m}$ for each $l$ and $m$ in light of the exchange symmetry.
The vertical black dashed lines and the vertical black dash-dotted ones respectively show the true value $b=b_0$ and the estimated value $b=\hat{b}$. The horizontal black dotted lines in (A) show the true value $\mu_k^*$ for each $k$ and the horizontal gray dashed line in (B) shows the spectral resolution $\Delta x$. The vertical black solid lines in (A-B) correspond to each value of $b$ in (a-d). The relative frequency histograms (a-d) show the marginal posterior probability of $\mu_k$ for each bin $[X_i, X_{i+1}]$ and $b$ as follows: 
\begin{linenomath*} 
\begin{align}
P(X_i \leq \mu_k \leq X_{i+1} \mid D, K, b) &= \int_{X_i}^{X_{i+1}} d\mu_k p(\mu_k \mid D, K, b), \\
p(\mu_k \mid D, K, b) &= \int dw' p(w \mid D, K, b) \\
&= \frac{\tilde{z}_n(K,b,\mu_k) \varphi(\mu_k)}{\tilde{Z}_n(K,b)}, \\
\tilde{z}_n(K,b,\mu_k) &:= \int dw' \exp \left[ -nb E_n(w'; \mu_k) \right] \varphi(w' \mid K),
\end{align}
\end{linenomath*} 
where $w' := w \backslash  \{ \mu_k \}$ and $\varphi(w' \mid K) := \varphi(w \mid K) / \varphi(\mu_k)$. $E_n(w'; \mu_k)$ indicates the function $E_n(w)$ given the value $\mu_k$.
The histograms (a), (b), and (d) were respectively constructed using the MCMC sample $\{\mu_{k,l,m}\}_{m=1}^{M_l}$ as $b=2.925210\times10^{-2}, 1.758132\times10^0, 6.350977\times10^3$ for each $k$.
Histogram (c) was calculated via Eq. (\ref{eq:reweighting}) for each $k$ (see Appendix C).
The coloring of the histogram for each $k$ follows that in (A-B). 
The horizontal axes in (a-d) represent $\mu_k$, and the vertical ones represent relative frequency on a log scale.
The vertical black dotted lines in (a-d) show the true value $\mu_k^*$ for each $k$, as in (A).
$\langle \mu_k \rangle_{b_l}$ and $2 \sqrt{\langle {\mu_k}^2 \rangle_{b_l} - {\langle \mu_k \rangle_{b_l}}^2}$ respectively change depending on $b$, where the changes in the support of the posterior density correspond.
These changes are considerable around $b=10^1$, where $\langle \mu_k \rangle_{b}$ for each $k$ asymptotically approaches the true value ${\mu_{k}}^*$ from this region and $2 \sqrt{\langle {\mu_k}^2 \rangle_{b} - {\langle \mu_k \rangle_{b}}^2}$ for each $k$ monotonically decreases from the same region. The marginal posterior densities of $\mu_1$, $\mu_2$, and $\mu_3$ overlap and are unidentifiable if $b$ is smaller than around $10^1$. Otherwise, they are separated and identifiable. 
$2 \sqrt{\langle {\mu_2}^2 \rangle_{b} - {\langle \mu_2 \rangle_{b}}^2}$ is smaller than $\Delta x$ as (c) $b=\hat{b}$: a kind of super-resolution.
This effect is based on the same principle as super-resolution microscopy techniques \cite{betzig1995proposed, betzig2006imaging}.
$2 \sqrt{\langle {\mu_k}^2 \rangle_{b} - {\langle \mu_k \rangle_{b}}^2}$ for each $k$ is also smaller than $\Delta x$ as (d) $b>\hat{b}$, whereas the support of $\mu_1$ does not cover the true value $\mu_1^*$: outside the confidence interval.
An appropriate setting of $b$ provides an appropriate precision of parameter estimation.
Estimating the optimal value of $b$ is indispensable even if the true model size $K_0$ is known; thus, this result also shows the validity of our framework.

\section{Discussion and Conclusion}
We constructed a framework that enables the dual estimation of the noise variance and the number of peaks and demonstrated the effectiveness of our framework through simulation.
We also warned that there are the risks of overfitting, overpenalizing, and misunderstanding the precision of parameter estimation without the estimation of the noise variance.
Our framework is an extension of Nagata et al.'s framework and is versatile and applicable to not only spectral deconvolution but also any other nonlinear regression with hierarchical statistical models.

Our framework is also considered as a learning scheme in radial basis function networks.
However, the goal of spectral deconvolution is not to predict any future data, which is the goal of most other learning tasks, but to identify the true model since spectral deconvolution is an inverse problem of physics.
This is the reason why we do not adopt the Bayes generalization error but adopt the Bayes free energy for hyperparameter optimization and model selection.
The Akaike information criterion (AIC) \cite{akaike1974new} and Bayesian information criterion (BIC) \cite{schwarz1978estimating}, which are respectively approximations of the generalization error and Bayes free energy, do not hold for hierarchical models such as radial basis function networks: the widely applicable information criterion (WAIC) \cite{watanabe2010equations} and widely applicable Bayesian information criterion (WBIC) \cite{watanabe2013widely} generally hold for any statistical model.
If the noise variance is unknown, these criteria do not lead to computational reduction since the value of the noise variance needs to be estimated, as discussed in Sect. \ref{sec:demo}. 
The example we gave is classified as an unrealizable and singular (or regular) case \cite{watanabe2010asymptotic}, which is a difficult problem.
On the other hand, the example Nagata et al. gave \cite{nagata2012bayesian} is classified as a realizable and singular (or regular) case, which is a relatively easy problem.
Statistical hypothesis testing does not hold for a singular case.
Our scheme is also valid and sophisticated from the viewpoint of statistics.  

\begin{acknowledgment}
This work was partially supported by a Grant-in-Aid for Scientific Research on Innovative Areas (No. 25120009) from the Japan Society for the Promotion of Science, by ``Materials Research by Information Integration'' Initiative (MI2I) project of the Support Program for Starting Up Innovation Hub from the Japan Science and Technology Agency (JST), and by Council for Science, Technology and Innovation (CSTI), Cross-ministerial Strategic Innovation Promotion Program (SIP), ``Structural Materials for Innovation'' (Funding agency: JST). 
\end{acknowledgment}

\section*{Appendix A: Bayes free energy for no-peaks model}
\renewcommand{\theequation}{A.\arabic{equation}}
\setcounter{equation}{0}

We define the function $f(x; w=\phi) = 0$ as $K=0$, where $\phi$ is the empty set.
The statistical model of the no-peaks spectrum and marginal likelihood are expressed as
\begin{linenomath*} 
\begin{align}
p(y \mid x, w=\phi, b) &= \sqrt{\frac{b}{2\pi}} \exp \left( - \frac{b}{2} y^2 \right), \\
Z_n(K=0,b) &= \prod_{i=1}^n p(Y_i \mid X_i, w=\phi,b) \\
&= \left( \frac{b}{2\pi} \right)^\frac{n}{2} \tilde{Z}_n(K=0,b), \\
\tilde{Z}_n(K=0,b) &= \exp[-nbE_n(w=\phi)], \\
E_n(w=\phi) &= \frac{1}{2n} \sum_{i=1}^n {Y_i}^2.
\end{align}
\end{linenomath*} 
The main term of Bayes free energy and the posterior mean of the mean square error are also respectively expressed as
\begin{linenomath*} 
\begin{align}
\tilde{F}_n(K=0,b) &= nE_n(w=\phi), \\
\langle E_n(w=\phi) \rangle_b &= E_n(w=\phi),
\end{align}
\end{linenomath*} 
where they can be calculated without any MCMC method.

\section*{Appendix B: Hierarchical Bayes approach}
\renewcommand{\theequation}{B.\arabic{equation}}
\setcounter{equation}{0}
\renewcommand{\thefigure}{B.\arabic{figure}}
\setcounter{figure}{0}

\begin{figure}[t]
\begin{tabular}{lll}
(A) & (B) & (C)\\
\vspace{-8pt}
\\
\includegraphics[width=4.75cm]{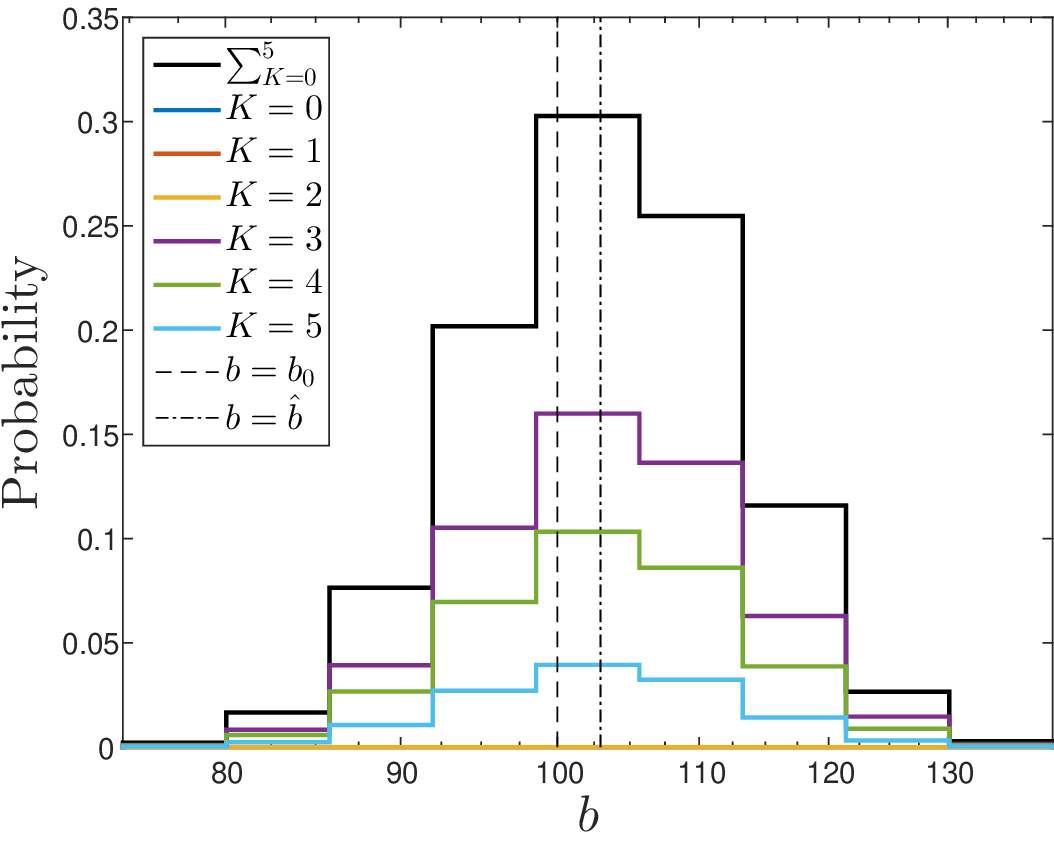}
&
\includegraphics[width=4.75cm]{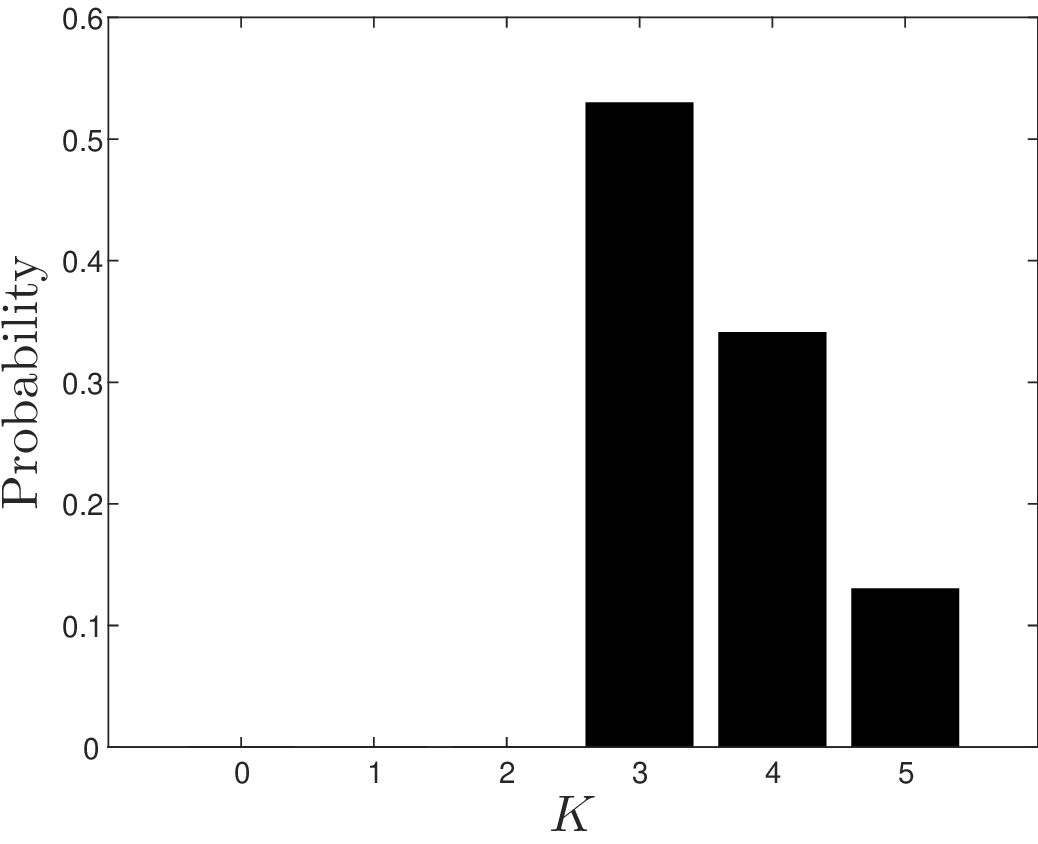}
&
\includegraphics[width=4.75cm]{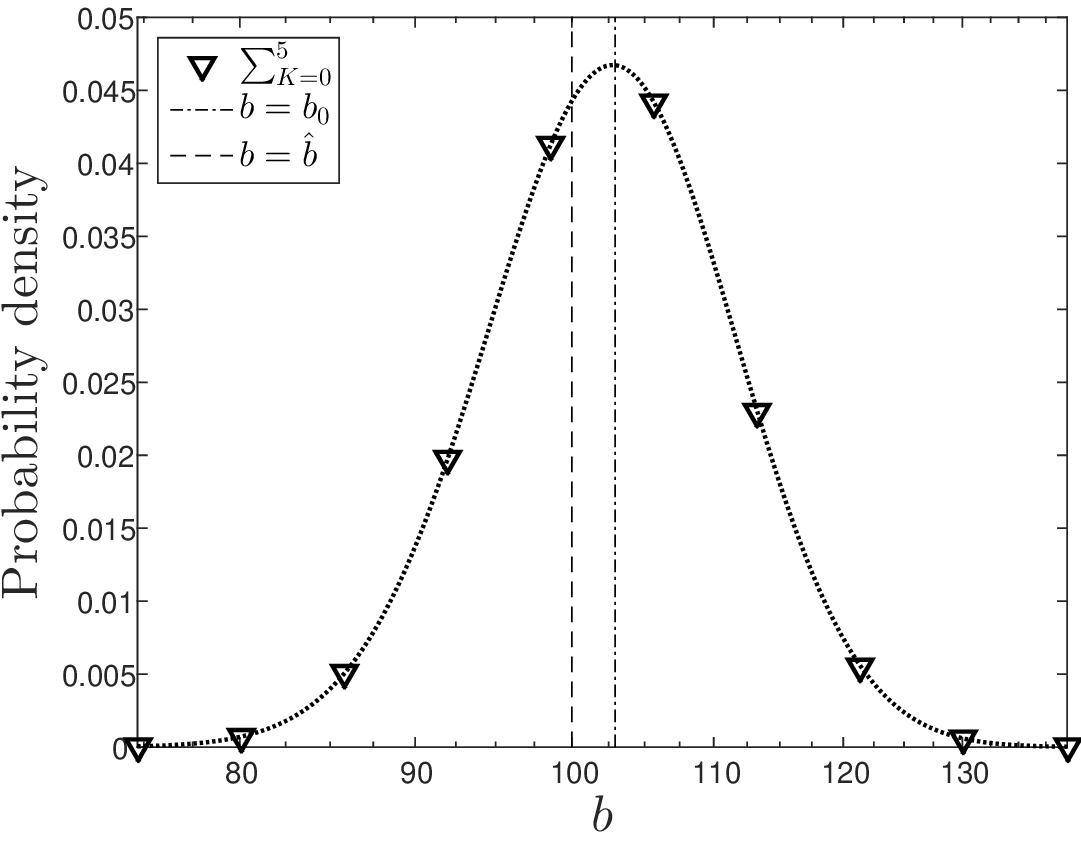}
\end{tabular}
\caption{(Color online) (A) Joint probability of $(K, b)$ and marginal probability of $b$, (B) marginal probability of $K$, and (C) marginal probability density of $b$. The horizontal axes represent $b$ on a log scale. The colored stairstep graphs and  the black one in (A) respectively show the joint probability $P(K, b_l \leq b \leq b_{l+1} \mid D)$ for each $K$ and the marginal probability $P(b_l \leq b \leq b_{l+1} \mid D)$. The three colored graphs of $K < 3$ almost overlap in contrast to Fig. \ref{fig:state_function}(A-1). The black bars in (B) show the marginal probability $P(K \mid D)$.
The black markers and black dotted line in (C) respectively show the marginal probability density $p(b_l \mid D)$ and the interpolated values. The vertical black dashed lines and the vertical black dash-dotted ones respectively show the true value $b=b_0$ and the estimated value $b=\hat{b}$, as in Fig. \ref{fig:state_function}. }
\label{fig:prob}
\end{figure}

In Sect. \ref{sec:demo}, we adopted the empirical Bayes (or type II maximum likelihood) approach, in which $K$ and $b$ are estimated by the minimization of $F_n(K,b)$ (or the maximization of $Z_n(K,b)$).
The hierarchical Bayes approach, which takes into account the posterior density of $K$ and $b$, is also suitable for our framework.
The prior density of $K$ and $b$ is set as $\varphi(K,b)=\varphi(K)\varphi(b)$, where $\varphi(K)$ is a discrete uniform distribution on the natural numbers $\{0, 1, 2, 3, 4, 5\}$ and $\varphi(b)$ is a continuous uniform distribution on the interval $[b_1, b_L]$.
The joint posterior probability and marginal ones are expressed as
\begin{linenomath*} 
\begin{align}
P(K, b_l \leq b \leq b_{l+1} \mid D) &= \int_{b_l}^{b_{l+1}} db p(K,b \mid D), \\
p(K, b_l \mid D) &= \frac{\exp[-F_n(K,b_l)]}{\sum_{K=0}^{5} \int_{b_1}^{b_{L}} db \exp[-F_n(K,b)]}, \\
P(K \mid D) &= \sum_{l=1}^{L-1} P(K, b_l \leq b \leq b_{l+1} \mid D), \\
P(b_l \leq b \leq b_{l+1} \mid D) &= \int_{b_l}^{b_{l+1}} db p(b \mid D), \\
p(b_l \mid D) &= \sum_{K=0}^{5} p(K,b_l \mid D), 
\end{align}
\end{linenomath*} 
where the integration along the $b$-axis is calculated using the trapezoidal rule.
Note that $\exp[-F_n(K,b_1)]=Z_n(K,b_1)=0$.
The (A) joint probability of $(K,b)$ and the marginal probability of $b$, (B) the marginal probability of $K$, and (C) the marginal probability density of $b$ are shown in Fig. \ref{fig:prob}. The horizontal axes represent $b$ on a log scale. The colored stairstep graphs and the black one in (A) respectively show the joint probability $P(K, b_l \leq b \leq b_{l+1} \mid D)$ for each $K$ and the marginal probability $P(b_l \leq b \leq b_{l+1} \mid D)$. 
The three colored graphs of $K < 3$ almost overlap in contrast to Fig. \ref{fig:state_function}(A-1).
The black bar in (B) shows the marginal probability $P(K \mid D)$.
The black markers and black dotted line in (C) respectively show the marginal probability density $p(b_l \mid D)$ and the interpolated values. The vertical black dashed lines and vertical black dash-dotted ones respectively show the true value $b=b_0$ and the estimated value $b=\hat{b}$, as in Fig. \ref{fig:state_function}.
Both $b_0$ and $\hat{b}$ are within the same interval of $b$, which maximize the probabilities $P(K, b_l \leq b \leq b_{l+1} \mid D)$ and $P(b_l \leq b \leq b_{l+1} \mid D)$ in this case.
Although the value of $K$ that maximizes $P(K \mid D)$ is the same as $\hat{K}$ in this case, the value of $b$ that maximizes $p(b \mid D)$ is slightly different from $\hat{b}$ in the strict sense.
These values are not always consistent in practice, and there is a continuous discussion:  which is better, to optimize or to integrate out? \cite{mackay1996hyperparameters}
The users of our framework can choose a better way in light of their perspective.

\section*{Appendix C: Interpolation of posterior distribution}
\renewcommand{\theequation}{C.\arabic{equation}}
\setcounter{equation}{0}

The density of states in the $i^{\rm th}$ bin, which is the function $g(E;K)$ given the value of $\mu_k$ in the interval $[X_i, X_{i+1}]$, is defined and estimated as
\begin{linenomath*} 
\begin{align}
g(E; K, X_i \leq \mu_k \leq X_{i+1}) &:= \int dw' \delta [E-E_n(w'; X_i \leq \mu_k \leq X_{i+1})] \varphi (w' \mid K) \\
&= \frac{\sum_{l=1}^L N_{l}(E; X_i \leq \mu_k \leq X_{i+1})}{\sum_{l'=1}^L M_{l'}^{(i)} \tilde{Z}_n(K,b_{l'})^{-1} \exp(-n b_{l'} E)},
\end{align}
\end{linenomath*} 
then we obtain
\begin{linenomath*} 
\begin{align}
\tilde{z}_n(K,b, X_i \leq \mu_k \leq X_{i+1}) &= \int dE g(E; K, X_i \leq \mu_k \leq X_{i+1}) \exp (-nbE) \\
&= \sum_{l=1}^L \sum_{m=1}^{M_{l}^{(i)}} \frac{1}{\sum_{l'=1}^L M_{l'}^{(i)} \tilde{z}_n(K,b_{l'}, X_i \leq \mu_k \leq X_{i+1})^{-1} \exp \left[ n (b-b_{l'}) E_{l, m}^{(i)} \right]}, \label{eq:small_marginal}
\end{align}
\end{linenomath*} 
where $E_n(w'; X_i \leq \mu_k \leq X_{i+1})$, $N_{l}(E; X_i \leq \mu_k \leq X_{i+1})$, and $E_{l, m}^{(i)}$ respectively indicate $E_n(w)$, $N_l(E)$, and $E_{l,m}$ in the $i^{\rm th}$ bin.
$M_{l}^{(i)}$ is defined as $M_{l}^{(i)} :=\int dE N_l(E; X_i \leq \mu_k \leq X_{i+1})$, where $M_{l} = \sum_{i=1}^{n-1} M_{l}^{(i)}$. The values of $\{\tilde{z}_n(K,b_l, X_i \leq \mu_k \leq X_{i+1})\}_{l=1}^L$ for each $i$ are determined self-consistently by iterating Eq. (\ref{eq:small_marginal}) with $b=b_l$.
Given $\{\tilde{z}_n(K,b_l, X_i \leq \mu_k \leq X_{i+1})\}_{l=1}^L$ for each $i$, we calculate $\tilde{z}_n(K,b, X_i \leq \mu_k \leq X_{i+1})$ for each $i$ with $b=b'$ via Eq. (\ref{eq:small_marginal}) again. If $\Delta x$ is sufficiently small (or $\varphi(\mu_k)$ is almost flat), $P(X_i \leq \mu_k \leq X_{i+1} \mid D, K, b)$ is expressed as
\begin{linenomath*} 
\begin{align}
P(X_i \leq \mu_k \leq X_{i+1} \mid D, K, b) 
&= \frac{\tilde{z}_n(K,b, X_i \leq \mu_k \leq X_{i+1}) \varphi(\mu_k=X_i)}{\sum_{i=1}^{n} \tilde{z}_n(K,b, X_i \leq \mu_k \leq X_{i+1}) \varphi(\mu_k=X_i)}. \label{eq:reweighting} 
\end{align}
\end{linenomath*} 

\bibliographystyle{jpsj}
\bibliography{reference}

\begin{thebibliography}{10}

\bibitem{tkachenko2006optical}
N.~V. Tkachenko: {\em Optical Spectroscopy: Methods and Instrumentations}
  (Elsevier, 2006).

\bibitem{allen1978deconvolution}
G.~C. Allen and R.~F. McMeeking: Anal. Chim. Acta {\bfseries 103} (1978) 73.

\bibitem{fischer2001analysis}
R.~Fischer and V.~Dose: Bayesian Methods: With Applications to Science, Policy,
  and Official Statistics, 2001, pp. 145--154.

\bibitem{razul2003bayesian}
S.~G. Razul, W.~Fitzgerald, and C.~Andrieu: Nucl. Instr. Meth. Phys. Res.
  {\bfseries 497} (2003) 492.

\bibitem{masson2010dynamics}
A.~Masson, L.~Poisson, M.-A. Gaveau, B.~Soep, J.-M. Mestdagh, V.~Mazet, and
  F.~Spiegelman: J. Chem. Phys {\bfseries 133} (2010) 054307.

\bibitem{nagata2012bayesian}
K.~Nagata, S.~Sugita, and M.~Okada: Neural Netw. {\bfseries 28} (2012) 82.

\bibitem{mazet2015unsupervised}
V.~Mazet, S.~Faisan, S.~Awali, M.-A. Gaveau, and L.~Poisson: Signal Processing
  {\bfseries 109} (2015) 193.

\bibitem{kasai2016nmr}
T.~Kasai, K.~Nagata, M.~Okada, and T.~Kigawa: JPCS, Vol. 699, 2016, p. 012003.

\bibitem{hong2016automatic}
P.~Hong, H.~Miyamoto, T.~Niihara, S.~Sugita, K.~Nagata, J.~M. Dohm, and
  M.~Okada: J. Geol. Geop. {\bfseries 5} (2016) 2.

\bibitem{PhysRevC.93.061601}
K.~Hagino: Phys. Rev. C {\bfseries 93} (2016) 061601.

\bibitem{murata2016extraction}
S.~Murata, K.~Nagata, M.~Uemura, and M.~Okada: J. Phys. Soc. Jpn {\bfseries 85}
  (2016) 104003.

\bibitem{hukushima1996exchange}
K.~Hukushima and K.~Nemoto: J. Phys. Soc. Jpn {\bfseries 65} (1996) 1604.

\bibitem{geyer1991markov}
C.~J. Geyer: Proc. of the 23rd Symposium on the Interface, 1991, pp. 156--163.

\bibitem{green1995reversible}
P.~J. Green: Biometrika {\bfseries 82} (1995) 711.

\bibitem{jasra2007population}
A.~Jasra, D.~A. Stephens, and C.~C. Holmes: Biometrika {\bfseries 94} (2007)
  787.

\bibitem{efron1973stein}
B.~Efron and C.~Morris: JASA {\bfseries 68} (1973) 117.

\bibitem{akaike1980likelihood}
H.~Akaike: Trabajos de estad{\'\i}stica y de investigaci{\'o}n operativa
  {\bfseries 31} (1980) 143.

\bibitem{mackay1992bayesian}
D.~J. MacKay: Neural Comput. {\bfseries 4} (1992) 415.

\bibitem{ferrenberg1989optimized}
A.~M. Ferrenberg and R.~H. Swendsen: Phys. Rev. Lett. {\bfseries 63} (1989)
  1195.

\bibitem{kumar1992weighted}
S.~Kumar, J.~M. Rosenberg, D.~Bouzida, R.~H. Swendsen, and P.~A. Kollman: J.
  Comput. Chem. {\bfseries 13} (1992) 1011.

\bibitem{loudon2000quantum}
R.~Loudon: {\em The Quantum Theory of Light} (Oxford University Press, 2000).

\bibitem{broomhead1988radial}
D.~S. Broomhead and D.~Lowe: Complex Systems {\bfseries 2} (1988) 321.

\bibitem{tokuda2013numerical}
S.~Tokuda, K.~Nagata, and M.~Okada: IPSJ TOM {\bfseries 6} (2013) 117.

\bibitem{gelman2013bayesian}
A.~Gelman, J.~Carlin, H.~Stern, D.~Dunson, A.~Vehtari, and D.~Rubin: {\em
  Bayesian Data Analysis, 3rd ed} (CRC Press, 2013).

\bibitem{metropolis1953equation}
N.~Metropolis, A.~W. Rosenbluth, M.~N. Rosenbluth, A.~H. Teller, and E.~Teller:
  J. Chem. Phys {\bfseries 21} (1953) 1087.

\bibitem{meng1996simulating}
X.-L. Meng and W.~H. Wong: Statistica Sinica  (1996) 831.

\bibitem{gelman1998simulating}
A.~Gelman and X.-L. Meng: Statistical Science  (1998) 163.

\bibitem{newman1999monte}
M.~Newman and G.~Barkema: {\em Monte Carlo Methods in Statistical Physics}
  (Oxford University Press, 1999).

\bibitem{watanabe2001algebraic}
S.~Watanabe: Neural Comput. {\bfseries 13} (2001) 899.

\bibitem{watanabe2009algebraic}
S.~Watanabe: {\em Algebraic Geometry and Statistical Learning Theory}
  (Cambridge University Press, 2009), Vol.~25.

\bibitem{betzig1995proposed}
E.~Betzig: Opt. Lett. {\bfseries 20} (1995) 237.

\bibitem{betzig2006imaging}
E.~Betzig, G.~H. Patterson, R.~Sougrat, O.~W. Lindwasser, S.~Olenych, J.~S.
  Bonifacino, M.~W. Davidson, J.~Lippincott-Schwartz, and H.~F. Hess: Science
  {\bfseries 313} (2006) 1642.

\bibitem{akaike1974new}
H.~Akaike: IEEE Trans. Automat. Contr. {\bfseries 19} (1974) 716.

\bibitem{schwarz1978estimating}
G.~Schwarz: Ann. Stat. {\bfseries 6} (1978) 461.

\bibitem{watanabe2010equations}
S.~Watanabe: Neural Netw. {\bfseries 23} (2010) 20.

\bibitem{watanabe2013widely}
S.~Watanabe: J. Mach. Learn. Res. {\bfseries 14} (2013) 867.

\bibitem{watanabe2010asymptotic}
S.~Watanabe: JPCS, Vol. 233, 2010, p. 012014.

\bibitem{mackay1996hyperparameters}
D.~J. MacKay: Maximum Entropy and Bayesian Methods, 1996, pp. 43--59.

\end{thebibliography}

\end{document}